\input{aipcheck}

\documentclass[
  ,final            
  ]
  {aipproc}
\layoutstyle{6x9}

\newcommand{\ud}{\mathrm{d}}

\begin{document}

\title{Quark Wigner Distributions}


\author{B.~Pasquini}{
  address={Dipartimento di Fisica Nucleare e Teorica, Universit\`a degli Studi di Pavia, \\and INFN, Sezione di Pavia, I-27100 Pavia, Italy}
}

\author{C. Lorc\'e}{
  address={Institut f\"ur Kernphysik, Johannes Gutenberg-Universit\"at,\\ D-55099 Mainz, Germany}
}

\begin{abstract}
We study the  Wigner functions of the nucleon  which provide multidimensional images of the quark distributions in phase space and combine in a single picture all the information contained in the generalized parton distributions (GPDs) and the transverse-momentum dependent parton distributions (TMDs). In particular, we present results for the distribution of unpolarized quark in a longitudinally polarized nucleon. Treating this distribution as it was a classical distribution, we also obtain the expectation value of the orbital angular momentum operator and compare the corresponding results from different quark models with the values obtained using alternative definitions of the orbital angular momentum, as given in terms of the GPDs and the TMDs.
\end{abstract}

\maketitle

\section{Introduction}
\label{section-1}

Quark Wigner distributions open a new way to map the distributions of momentum and spin of the proton onto its constituents. They provide joint position-and-momentum (or phase-space) distributions, encoding in a unified picture  the information  obtained from transverse-momentum dependent parton distributions (TMDs) and generalized parton distributions (GPDs) in impact-parameter space. The concept of Wigner distributions in QCD for quarks and gluons was first explored in Refs.~\cite{Ji:2003ak,Belitsky:2003nz}. Neglecting relativistic effects, the authors introduced six-dimensional (three position and three momentum coordinates) Wigner distributions. In a recent work~\cite{Lorce:2011kd}, we used the connection between Wigner distributions and generalized transverse-momentum dependent parton distributions (GTMDs)~\cite{Meissner:2009ww} to study five-dimensional distributions (two position and three momentum coordinates) as seen from the infinite-momentum frame (IMF). The main advantage of working in the IMF is that one obtains  Wigner distributions which are completely consistent with special relativity. However, it is well known that the phase-space distributions do not have a density interpretation because the uncertainty principle prevents to determine simultaneously position and momentum of a quantum-mechanical system. Accordingly, Wigner distributions are not positively defined. Nevertheless, the physics of the Wigner distributions is very rich and one can select certain situations where a semiclassical interpretation is still possible.

The purpose of this contribution is to investigate the phenomenology of the quark Wigner distributions. As a matter of fact, since it is not known how to access these distributions directly from experiments, phenomenological models are very powerful in this context. Collecting the information that one can learn from quark models which were built up on the basis of available experimental information on GPDs and TMDs, one can hope to reconstruct a faithful description of the physics of the Wigner distributions. To this aim we will rely on models for the light-cone wave functions which have already been used for the description of the GPDs~\cite{Boffi:2007yc,Lorce:2011dv}, the TMDs~\cite{Lorce:2011dv,Pasquini:2008ax,Boffi:2009sh,Pasquini:2010af,Lorce:2011zt} and  electroweak properties of the nucleon~\cite{Lorce:2011dv,Lorce:2006nq,Lorce:2007as,Lorce:2007fa,Pasquini:2007iz}.

\section{Wigner Operators and Wigner Distributions}
\label{section-2a}

Similarly to Refs.~\cite{Ji:2003ak,Belitsky:2003nz}, we define the Hermitian Wigner operators for quarks at a fixed light-cone time $y^+=0$ as follows
\begin{equation}\label{wigner-operator}
\widehat W^{[\Gamma]}(\vec b_\perp,\vec k_\perp,x)\equiv\frac{1}{2}\int\frac{\ud z^-\,\ud^2z_\perp}{(2\pi)^3}\,e^{i(xp^+z^--\vec k_\perp\cdot\vec z_\perp)}\,\overline{\psi}\left(y-\frac{z}{2}\right)\Gamma\mathcal W\,\psi\left(y+\frac{z}{2}\right)\big|_{z^+=0}
\end{equation}
with $y^\mu=[0,0,\vec b_\perp]$, $p^+$ the average nucleon longitudinal momentum and $x=k^+/p^+$ the average fraction of nucleon longitudinal momentum carried by the active quark. The superscript $\Gamma$ stands for any twist-two Dirac operator $\Gamma=\gamma^+,\gamma^+\gamma_5,i\sigma^{j+}\gamma_5$ with $j=1,2$. A Wilson line $\mathcal W\equiv\mathcal W(y-\frac{z}{2},y+\frac{z}{2}|n)$ ensures the color gauge invariance of the Wigner operator. In the following we will focus on  the quark contribution, ignoring the gauge-field degrees of freedom and therefore reducing the gauge link $\mathcal W$ to the identity.

We define the Wigner distributions $\rho^{[\Gamma]}(\vec b_\perp, \vec k_\perp,x,\vec S)$ in terms of the matrix elements of the Wigner operators~(\ref{wigner-operator}) sandwiched between nucleon states with polarization $\vec S$. As outlined in Ref.~\cite{Lorce:2011kd}, such matrix elements can easily be interpreted as two-dimensional Fourier transforms of the GTMDs in the impact parameter space. Although the GTMDs are in general complex-valued functions, their two-dimensional Fourier transforms are always real-valued functions, in accordance with their interpretation as phase-space distributions. We note that $\vec b_\perp$ and $\vec k_\perp$ are not Fourier conjugate variables, like in the usual quantum-mechanical Wigner distributions. However, they are subjected to Heisenberg's uncertainty principle because the corresponding quantum-mechanical operators do not commute $[\hat{\vec b}_\perp,\hat{\vec k}_\perp]\neq 0$. As a consequence, the Wigner functions can not have a strict probabilistic interpretation. There are in total 16 Wigner functions at twist-two level, corresponding to all the 16 possible configurations of nucleon and quark polarizations. Here we will discuss only one particular case, namely the distortion of the distribution of unpolarized quarks due to the longitudinal polarization of the nucleon $\rho_{LU}=\rho^{[\gamma^+]}(\vec b_\perp, \vec k_\perp,x,+\vec e_z)-\rho^{[\gamma^+]}(\vec b_\perp, \vec k_\perp,x,-\vec e_z)$.

\begin{figure}[t!]
	\centering
		\includegraphics[width=.49\textwidth]{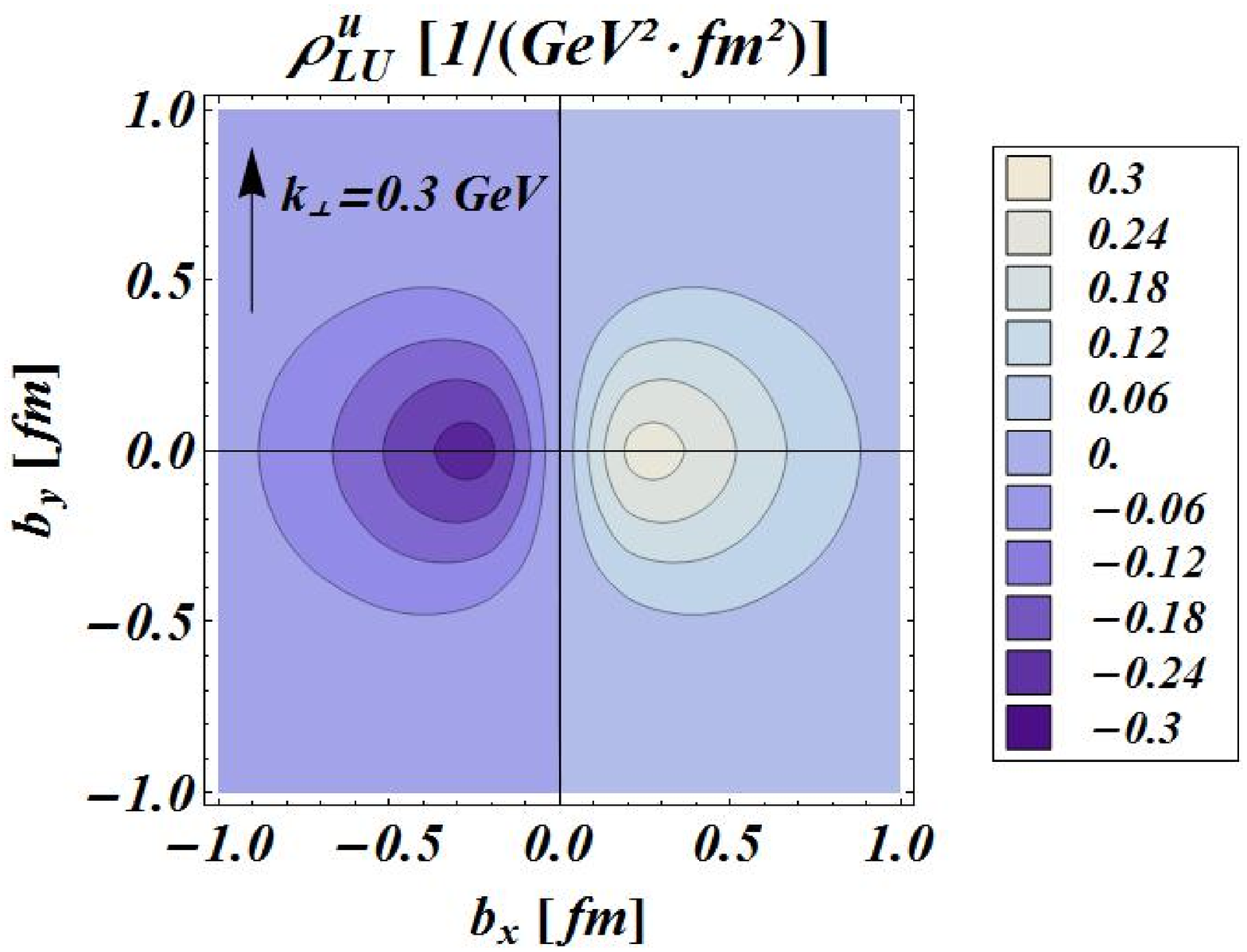}
		\includegraphics[width=.49\textwidth]{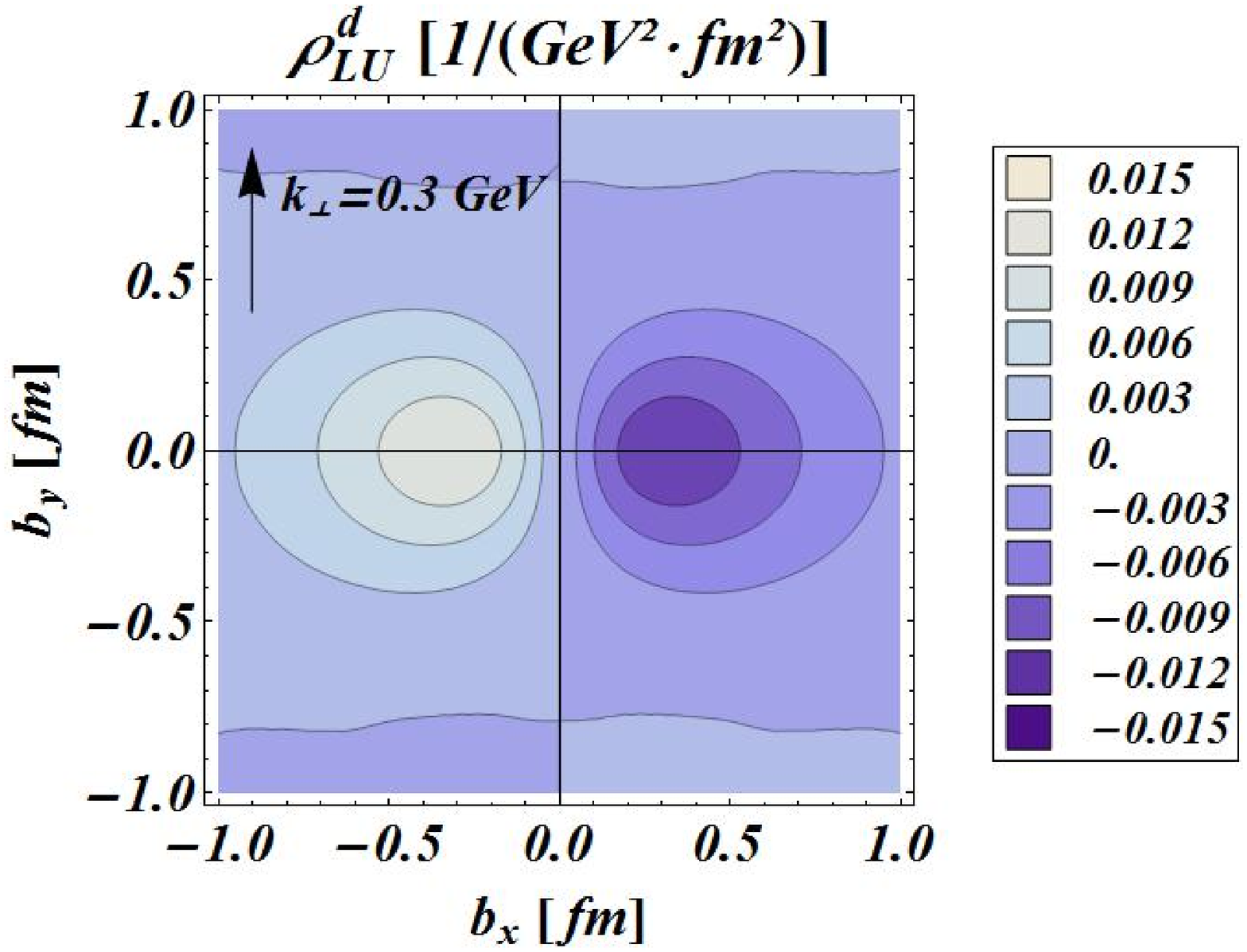}
\end{figure}
\begin{figure}[th!]
	\centering
		\includegraphics[width=.49\textwidth]{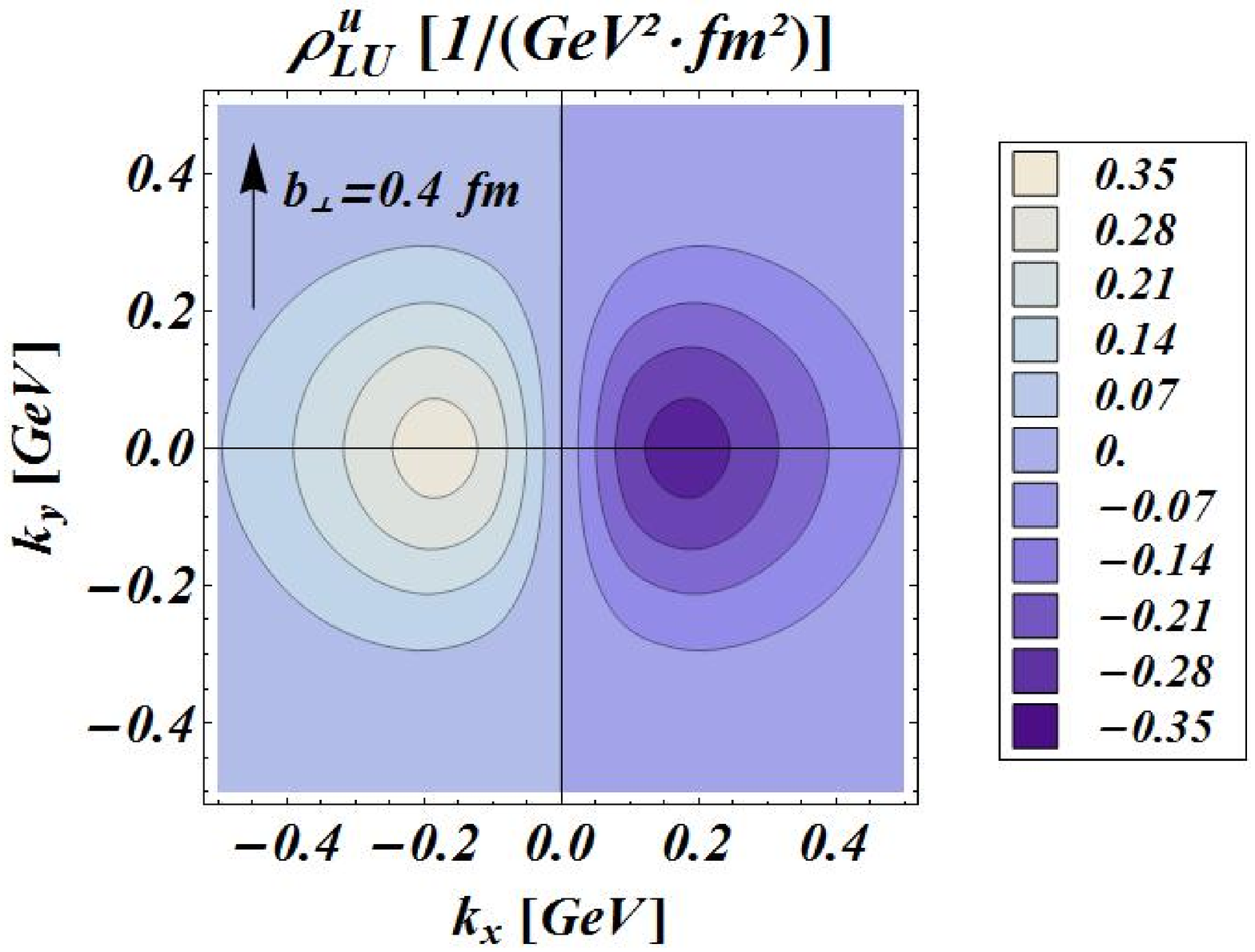}
		\includegraphics[width=.49\textwidth]{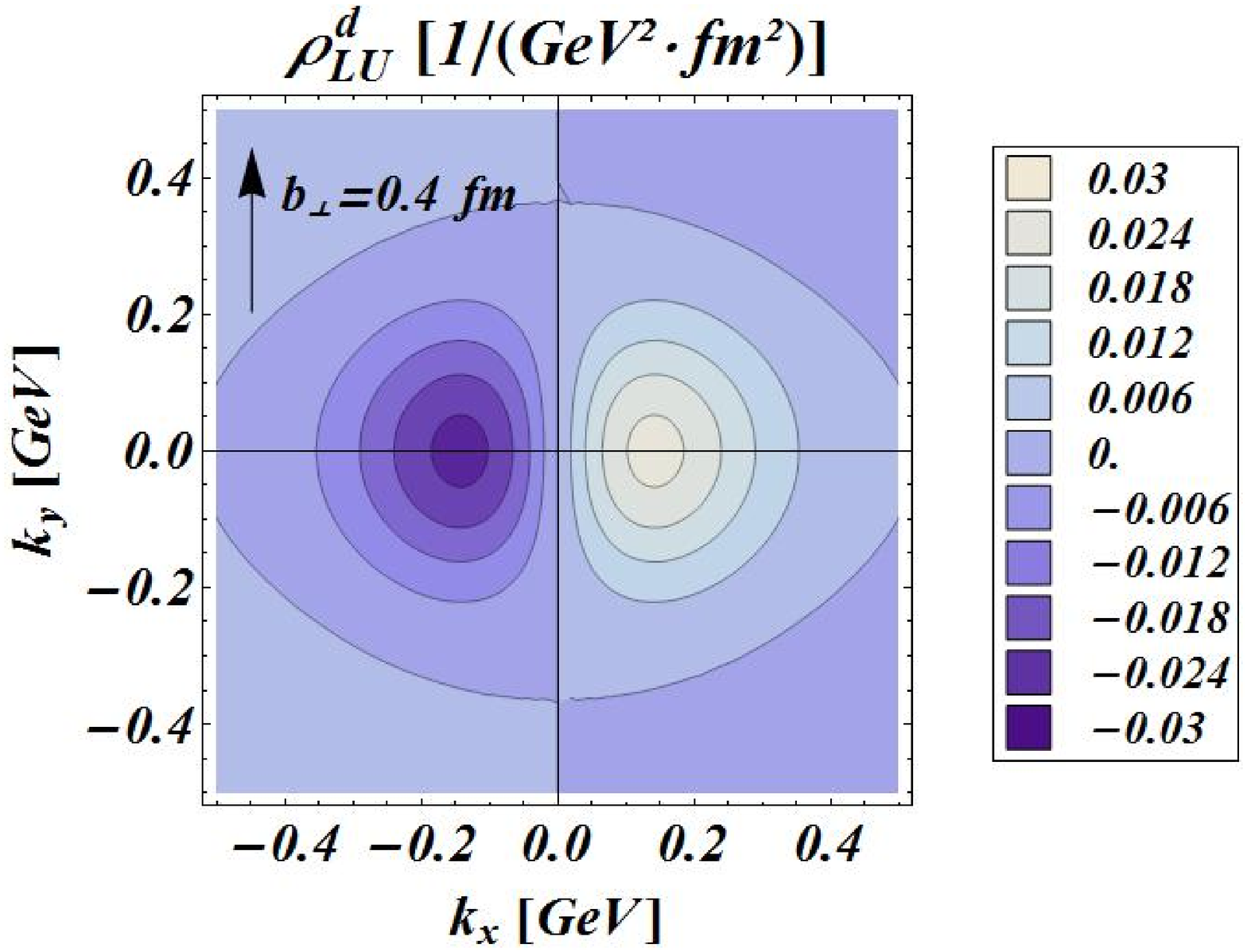}
		\caption{\footnotesize{The distortions of the transverse Wigner distributions of unpolarized quarks due to the spin of the proton (pointing out of the plane). Upper panels:  distortions in impact-parameter space with fixed transverse momentum $\vec k_\perp=k_\perp\,\hat e_y$ and $k_\perp=0.3$ GeV. Lower panels: distortions in transverse-momentum space with fixed impact parameter $\vec b_\perp=b_\perp\,\hat e_y$ and $b_\perp=0.4$ fm. The left (right) panels show the results for $u$ ($d$) quarks.}}\label{fig4}
\end{figure}
In Fig.~\ref{fig4}, the upper panels show the distortions in impact-parameter space for $u$ (left panels) and $d$ (right panels) quarks with fixed transverse momentum $\vec k_\perp=k_\perp\,\hat e_y$ and $k_\perp=0.3$ GeV, while the lower panels give the corresponding distortions in the transverse-momentum space with fixed impact parameter $\vec b_\perp=b_\perp \hat e_y$ and $b_\perp=0.4 $ fm.

We observe a clear dipole structure in both these distributions, with opposite sign for $u$ and $d$ quarks. We learn from these figures that the orbital angular momentum (OAM) of $u$ quarks tends be aligned with the nucleon spin, while the OAM of $d$ quarks tends be antialigned with the nucleon spin. In particular, we notice that the distortion induced by the quark OAM is stronger in the central region of the phase space ($k_\perp\ll$ and $b_\perp\ll$), for both $u$ and $d$ quarks. The distortion in the $\vec b_\perp$ space (see upper panels of Fig.~\ref{fig4}) is more extended for $d$ quarks than for $u$ quarks, whereas the opposite behavior is found for the distortion in the $\vec k_\perp$ space (see lower panels of Fig.~\ref{fig4}). In the case of $d$ quarks, we also observe a sign change of the distributions in the outer regions of phase space ($k_\perp\gg$ and $b_\perp\gg$) which corresponds to a flip of the local net quark OAM. By treating the Wigner functions as if they were classical distributions, we can obtain the expectation value of the quark OAM operator by calculating the integral over the phase space of the distribution in Fig.~\ref{fig4} multiplied by $(\vec b_\perp\times \vec k_\perp)_z$. This definition differs from the quark OAM calculated from GPDs, using the Ji's sum rule~\cite{Ji:1996ek}, as well as from the quark OAM defined in terms of the $h_{1T}^\perp$ TMD~\cite{She:2009jq,Avakian:2010br}. However, it turns out that in models without gauge-field degrees of freedom all  the three definitions give the same values for the total quark contributions, while they differ for the individual $u$ and $d$ contributions. In particular, we present in Table~\ref{OAMtable} the results from the LCCQM and the   light-cone version of the chiral quark-soliton model ($\chi$QSM) restricted to the three-quark sector \cite{Lorce:2006nq,Lorce:2007as,Lorce:2007fa}.
\begin{table}[th!]
\begin{tabular}{c|ccc|ccc}
\hline
Model
&\multicolumn{3}{c|}{LCCQM}
&\multicolumn{3}{c}{$\chi$QSM}\\
q
&$u$&$d$ &TOT&$u$&$d$ &TOT\\
\hline
$\ell^q_z$&$0.131$&$-0.005$&$0.126$&$0.073$&$-0.004$&$0.069$
\\
$L^q_z$&$0.071$&$0.055$&$0.126$&$-0.008$&$0.077$&$0.069$\\
$\mathcal L^q_z$ & $0.169$ &$-0.042$ &$0.126$ &$0.093$
&$-0.023$&$0.069$
\\
\hline
\end{tabular}
\caption{
\footnotesize{The results for quark OAM obtained from the Wigner functions ($\ell^q_z$), from the Ji's sum rule ($L_z$), and from the $h_{1T}^\perp$ TMD ($\mathcal L^q_z$) within the  LCCQM and the $\chi$QSM for $u$-, $d$- and total ($u+d$) quark contributions.}}\label{OAMtable}
\end{table}
In the LCCQM, there is more net quark OAM ($\sum_q L^q_z=0.126$) than in the $\chi$QSM ($\sum_q L^q_z=0.069$). For the individual quark contributions, both the LCCQM and the $\chi$QSM predict that $\ell^q_{z}$ (from the Wigner functions) and  $\mathcal L^q_z$  (from the $h_{1T}^\perp$ TMD) are positive for $u$ quarks and negative for $d$ quarks, with the $u$-quark contribution larger than the $d$-quark contribution in absolute value. For $L_z^q$ (from the Ji's sum rule) the LCCQM predicts the same positive sign for the $u$ and $d$ contributions, with the isovector combination $L^u_z- L^d_z>0$, similarly to a variety of relativistic quark model calculations. Instead, the $\chi$QSM gives $L_z^u<0$  and $L_z^d>0$, and therefore $L^u_z-L^d_z<0$, in agreement with lattice calculations~\cite{Hagler:2007xi}.

In summary, we have presented the first model calculation of the Wigner distributions within the light-cone formalism, using a derivation which is not spoiled by relativistic corrections. The results  within a light-cone constituent quark model and the light-cone version of the chiral quark-soliton model are very similar, and allowed us to sketch some general features about the behavior of the quarks in the nucleon when observed in the $\vec b_\perp$ plane at fixed $\vec k_\perp$, or in the $\vec k_\perp$ plane at fixed $\vec b_\perp$.





\bibliographystyle{aipproc}   

\bibliography{mybiblio}

\IfFileExists{\jobname.bbl}{}
 {\typeout{}
  \typeout{******************************************}
  \typeout{** Please run "bibtex \jobname" to optain}
  \typeout{** the bibliography and then re-run LaTeX}
  \typeout{** twice to fix the references!}
  \typeout{******************************************}
  \typeout{}
 }

\end{document}